# Evaluating the Efficacy of ChatGPT-4 in Providing Scientific References across Diverse Disciplines

Zhi Cao[1]

**Abstract**

This work conducts a comprehensive exploration into the proficiency of OpenAI's ChatGPT-4 in sourcing scientific references within an array of research disciplines. Our in-depth analysis encompasses a wide scope of fields including Computer Science (CS), Mechanical Engineering (ME), Electrical Engineering (EE), Biomedical Engineering (BME), and Medicine, as well as their more specialized sub-domains. Our empirical findings indicate a significant variance in ChatGPT-4's performance across these disciplines. Notably, the validity rate of suggested articles in CS, BME, and Medicine surpasses 65%, whereas in the realms of ME and EE, the model fails to verify any article as valid. Further, in the context of retrieving articles pertinent to niche research topics, ChatGPT-4 tends to yield references that align with the broader thematic areas as opposed to the narrowly defined topics of interest. This observed disparity underscores the pronounced variability in accuracy across diverse research fields, indicating the potential requirement for model refinement to enhance its functionality in academic research. Our investigation offers valuable insights into the current capacities and limitations of AI-powered tools in scholarly research, thereby emphasizing the indispensable role of human oversight and rigorous validation in leveraging such models for academic pursuits.

**Introduction**

The rise of artificial intelligence (AI) has led to significant breakthroughs across many fields, one notable area being language comprehension and generation. The rise of large language models (LLM), which was trained on copious amounts of data, including books, articles, websites, and other texts, allows computer to acquire a profound understanding of language structure and context.

OpenAI's ChatGPT-4, a recently developed large language model (LLM), has demonstrated its potential by successfully undergoing evaluations in diverse examinations[1-5], including the US Fundamentals of Engineering Exam[6], the Non-English National Medical Licensing Examination[7], and the ACR Radiation Oncology In-Training Exam[8]. While these achievements highlight ChatGPT-4's remarkable ability to handle complex exam-style questions across a wide array of domains, the data retention behavior of this model in providing succinct responses from its training dataset remains a subject of ongoing investigation.

Prior research has unveiled a significant correlation between the degree of memorization exhibited by LLMs and the frequency with which passages from the respective books appear on the web [9]. While this result demonstrates the capability of ChatGPT-4 in memorizing a wide collection of written materials, the specific ability of ChatGPT-4 in sourcing relevant written materials based on varying requirements remains a gap in the current body of knowledge.

[1] Email: Zhi.Cao001@umb.edu

In this study, we carry out a study to discover the ability of ChatGPT-4 in providing pertinent scientific articles across various research domains. Specifically, we design a systematic evaluation framework that testify the validity and relevance of the articles suggested by ChatGPT-4 across several broad fields including Computer Science (CS), Mechanical Engineering (ME), Electrical Engineering (EE), Biomedical Engineering (BME), and Medicine. Then, we delve deeper into these disciplines, evaluating the model's performance in sourcing articles related to more specialized sub-domains. Finally, we conclude our study with an analysis and discussion of the empirical findings, with a view to elucidating potential avenues for refining the capabilities of subsequent versions of AI models akin to ChatGPT-4.

## Experiment

The ChatGPT-4 under test is the March 23 version and in order to ensure a thorough assessment, we have selected five subjects: CS, ME, EE, BME, and Medicine. To glean a more comprehensive understanding of the model's proficiency across varied areas within these domains, we have chosen five unique research directions within each field for detailed investigation. The chosen research directions are summarized in Table 1.

| Subject | CS | ME | EE | BME | Medicine |
|---|---|---|---|---|---|
| Sub-direction | Computer Networking | Solid Mechanics | MEMS | Tissue Engineering | Immunology |
| | Computer Vision | Fluid Mechanics | Optics | Genetic Engineering | Computational Biology |
| | NLP | Robotics | Signal Processing | Biomaterial | Disease |
| | Cybersecurity | Biomechanics | Control System | Bioinformatics | Oncology |
| | Operating System | Manufacturing | Semiconductor | Neural Engineering | General Surgery |

Table 1. Disciplines and their respective subfields subjected to examination. NLP represents Natural Language Processing and MEMS stands for Micro-Electromechanical Systems.

Due to the absence of an API of GPT-4 at the time of completing this article, all processes and transactions were carried out within the chat interface. Figure 1 illustrates a representative example of how requests for scientific references in each research direction were initiated, and the answers were manually collected. For each request, we gathered information on 50 references. A two-step evaluation process was then employed for each reference obtained: firstly, we checked for the presence of the reference title using Google Scholar and verified the publication information, including authorship, journal, and year; the subsequent step entailed assessing the pertinence of the articles in relation to the research directions, accomplished by utilizing ChatGPT-4's capabilities to analyze their abstracts.

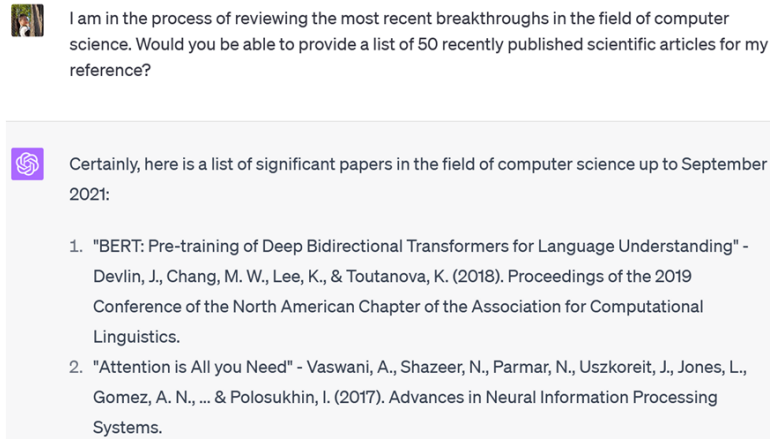

Figure 1. Typical chat interface layout

## Result

In order to evaluate the efficacy of ChatGPT-4 in provisioning scientific references spanning expansive research domains, including CS, ME, EE, BME, and Medicine, we undertook a systematic assessment. The result (Figure 2) exhibit that ChatGPT-4 demonstrated successful identification of legitimate articles in the domains of CS, BME, and Medicine, with an accuracy rate surpassing 65%. However, it is noteworthy that in the case of the ME and EE fields, the tool failed to verify any of the articles as valid. This observed discrepancy suggests a pronounced chasm in accuracy across different subject areas.

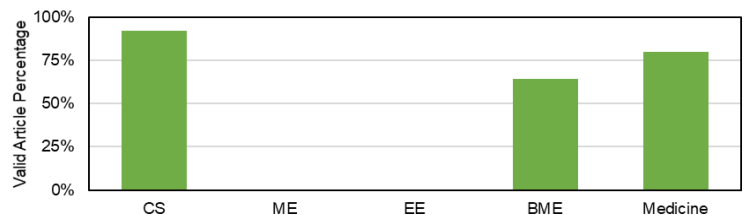

Figure 2. Proportion of verified articles within the reference list provided by ChatGPT-4 across varied research domains.

Analogous evaluations were then conducted on the sub-disciplines encompassed within each of these five subject areas. The results underscored palpable disparities in the veracity of the suggested articles across these sub-directions, even under the umbrella of the same overarching research subjects (Figure 3). For instance, under the subfield of NLP, 92% of the references were found to be authentic, complete with accurate publication information. In stark contrast, only approximately 45% of the references were validated under Computer Vision, and less than 10% were deemed authentic in the realms of Computer Networking, Cybersecurity and Operating Systems.

A similar pattern was observed in BME and Medicine, where significant deviations in the validity percentage were noted across various sub-disciplines. Intriguingly, almost none of the subfields within ME and EE were verified as authentic, barring Robotics, which had a validity quotient of 42%. These findings resonate with the previous results, further emphasizing a discernible variability in ChatGPT-4's performance across different sub-disciplines within the same broader scientific fields.

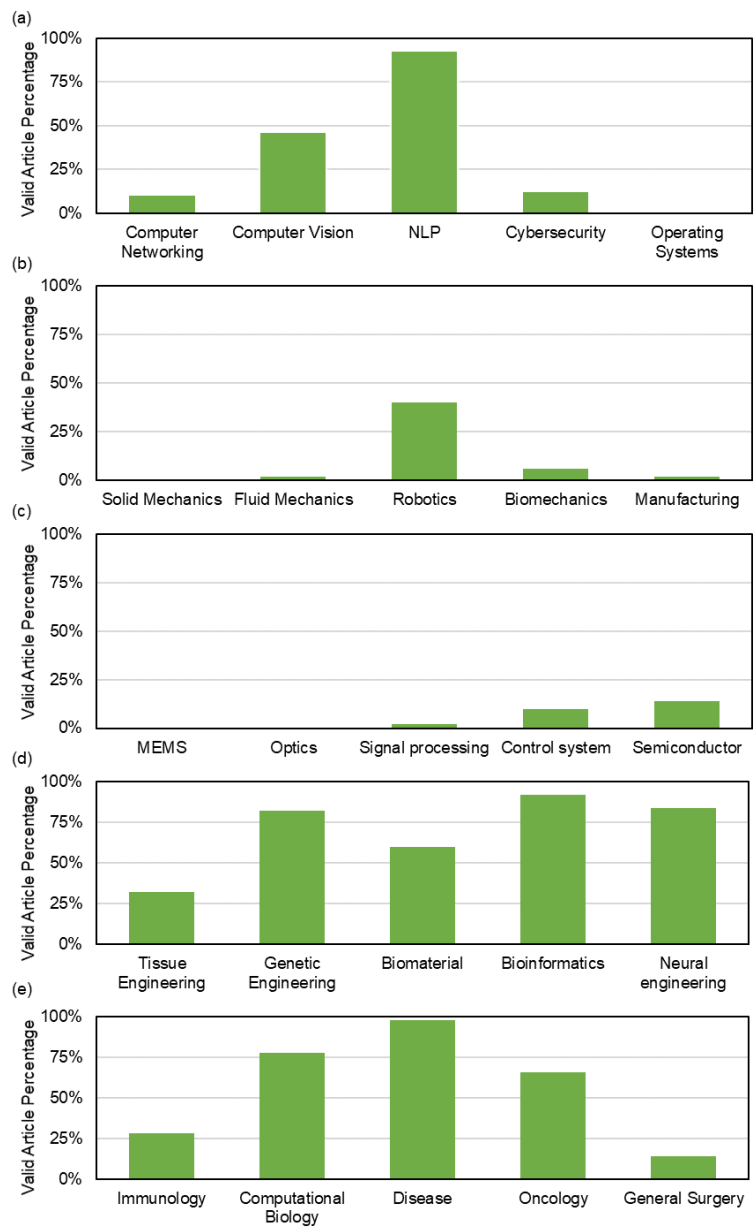

Figure 3. Proportion of verified articles within the reference list provided by ChatGPT-4 across varied research sub-directions under CS, ME, EE, BME, and Medicine.

In order to maintain uniformity in gauging the relevance of these papers to their respective research directions, abstracts of the articles were extracted from Google Scholar, conditional on their publication information being previously verified. These abstracts were then subjected to analysis by ChatGPT-4 to assess their correlation to the specific research directions. The resultant data, presented in Figure 4, portrays the distribution of relevance scores across the examined sub-directions. Noteworthy is the striking uniformity that unfolds, with a significant proportion of the articles yielding relevance scores surpassing the 70% mark across all sub-directions. This outcome underscores the pronounced alignment of the articles with their corresponding research directions.

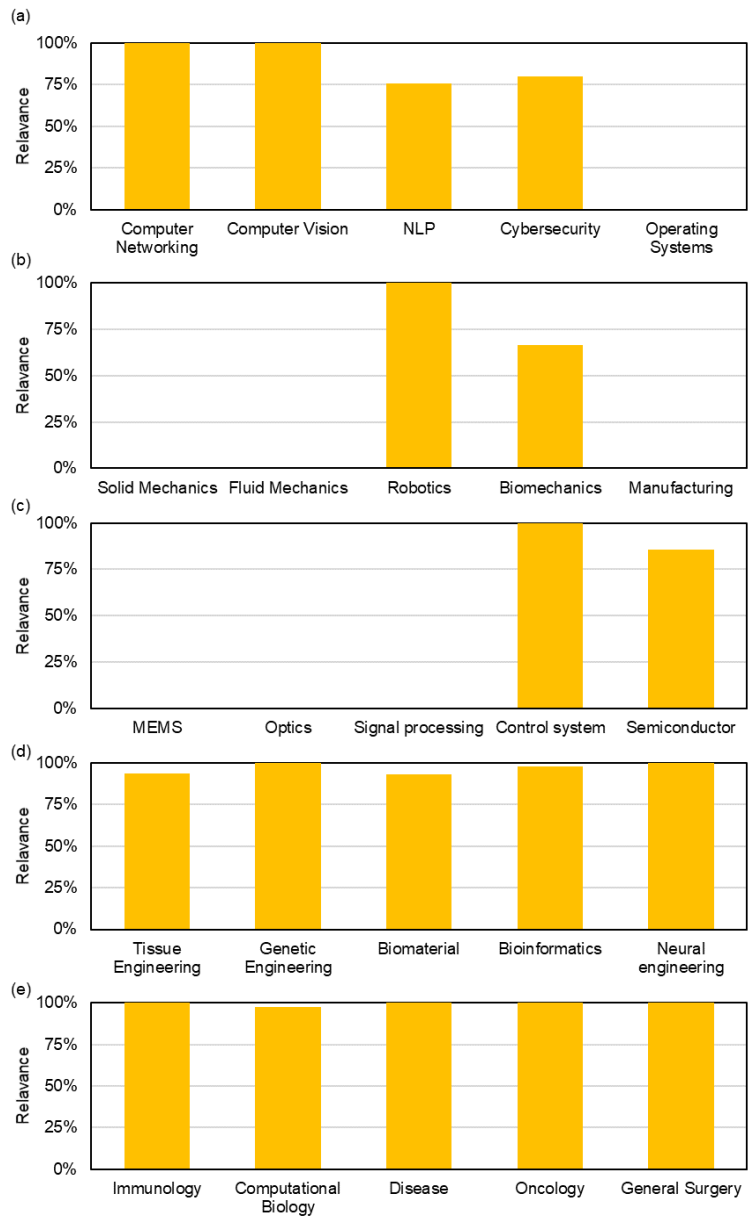

Figure 4. Relevance of validated scientific articles with specific research directions

It is important to underscore that these results may not fully encapsulate the underlying patterns given the relatively limited size of the sample sets utilized in the study. Despite this, they can still confer a preliminary insight into the performance of ChatGPT-4. Even with the inherent limitations, the data suggests that the references proposed by ChatGPT-4 do exhibit a satisfactory degree of alignment with the corresponding subfields, provided the identified articles pass the validity check.

In addition to assessing the overall relevance of the articles to their respective research directions, we also conducted focused tests to explore ChatGPT-4's performance in narrower research topics. We specifically examined two specific areas: Edge Computing within the domain of Computer Networking [10-11], and magnetohydrodynamic (MHD) nanofluid within Fluid Mechanics [12-13].The study found that none of the retrieved article titles were deemed applicable for MHD Nanofluid. This outcome could potentially indicate a limitation in the availability of pertinent research materials within this particular scientific area.

In contrast, Edge Computing demonstrated more promising results, with over 25% of the gathered references being authenticated (Figure 5). Yet, a more comprehensive scrutiny of the abstracts suggested that only a quarter of these publications were germane to the research field. When we refined the scope of our research direction further to task allocations [14-16], which is a popular research direction under the umbrella of Edge Computing, we still observed a validity rate exceeding 10% across all references. However, all these articles primarily dealt with general aspects of Edge Computing, instead of being dedicated to the specialized topic of task allocation.

These findings strongly hint towards ChatGPT-4's existing limitations in pinpointing relevant and valid articles for highly specialized research areas. This emphasizes the exigency for continual refinements and enhancements to this tool to ensure its optimality and effectiveness in various scientific research contexts.

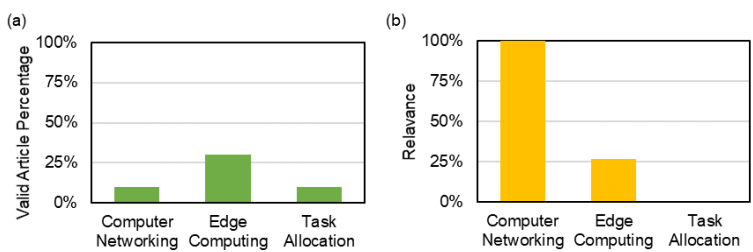

Figure 5. Percentage of valid articles pertaining to edge computing and task allocation and their corresponding relevance to these specific topics

## Discussion

Although the training dataset has never been disclosed, the mixed performance of ChatGPT-4 in retrieving scientific papers from different domains likely stems from the nature and structure of its training data. Given that the model is trained on a wide range of text from the internet, it stands to reason that the domains where it performs well, such as CS, BME, and Medicine, are likely to be areas where a larger quantity of easily accessible and comprehensible resources are available online. On the other hand, domains such as ME and EE, where ChatGPT-4 exhibited a less satisfactory performance, may have fewer resources available in the training data, or the resources may be more complex and less easily distilled into the model's training process.

The tendency of the model to prioritize broader themes when queried about specialized topics could be attributed to the general breadth-first approach of the model's training, where it learns broad, common patterns before it learns more specialized, narrower ones. This might also indicate a lack of sufficient training data regarding specific, specialized topics, thereby leading the model to default to broader, more general resources.

This highlights the potential for further research into the development of more specific and advanced training methods, which could include more specialized resources or incorporate a more deliberate focus on depth of understanding in particular domains. It also underscores the ongoing need for human intervention and validation in ensuring the precision and applicability of the model's outputs, particularly when it comes to specialized academic research.

## Conclusion

Our investigation into ChatGPT-4's proficiency in retrieving scientific papers presents a complex picture. The model exhibits commendable performance in certain domains like Computer Science, Biomedical Engineering, and Medicine, with the validation rate of the retrieved papers surpassing 65%. However, in Mechanical and Electrical Engineering domains, it fails to exhibit comparable success, signifying a considerable inconsistency in its performance across diverse areas. Additionally, when queried on highly specialized topics nested within broader domains, ChatGPT-4 tends to prioritize articles with broader themes over those precisely aligned with the given narrow topics. This tendency underscores a crucial limitation in ChatGPT-4's functionality in terms of scientific reference retrieval and emphasizes the need for advancements in its specificity and validity.

These findings suggest that while AI models such as ChatGPT-4 can be instrumental tools for initial research steps, the critical role of human validation and oversight cannot be understated, especially when it comes to ensuring the accuracy and relevance of the returned resources. Consequently, the evolution of such models should focus on amplifying their reliability across an array of topics and enhancing their capability to adapt to specificity levels as dictated by user requirements.